\documentclass{he_symp}
\usepackage{psfig,graphicx,epsfig}
\usepackage{color}
\usepackage{amsmath,amssymb,epic,eepic,array}
\begin{document}
\renewcommand{\FirstPageOfPaper }{ 114}\renewcommand{\LastPageOfPaper }{ 125}

\newcommand{\nodata}{$\cdot\cdot\cdot$}  
\newcommand{\de}{$^{\circ\ }$}
\newcommand{\Msolar}{{$\rm\ M_\odot$}}                         
\newcommand{\Mearth}{{$\rm\ M_\oplus$}}                             
\newcommand{\dddnu}{\stackrel{...}{\nu}}
\newcommand{\ddddnu}{\stackrel{....}{\nu}}

\let\tableline=\hline
\newcommand{\tablenotemarka}{$^{\rm a}$}
\newcommand{\tablenotemarkb}{$^{\rm b}$}
\newcommand{\tablenotemarkc}{$^{\rm c}$}
\newcommand{\tablenotemarkd}{$^{\rm d}$}
\newcommand{\tablenotemarke}{$^{\rm e}$}
\newcommand{\tablenotemarkf}{$^{\rm f}$}
\newcommand{\tablenotemarkg}{$^{\rm g}$}
\newcommand{\tablenotemarkh}{$^{\rm h}$}
\newcommand{\tablenotetexta}{\item[$^{\rm a}$]}
\newcommand{\tablenotetextb}{\item[$^{\rm b}$]}
\newcommand{\tablenotetextc}{\item[$^{\rm c}$]}
\newcommand{\tablenotetextd}{\item[$^{\rm d}$]}
\newcommand{\tablenotetexte}{\item[$^{\rm e}$]}
\newcommand{\tablenotetextf}{\item[$^{\rm f}$]}
\newcommand{\tablenotetextg}{\item[$^{\rm g}$]}
\newcommand{\tablenotetexth}{\item[$^{\rm h}$]}
\newcommand{\tablenotetexti}{\item[$^{\rm i}$]}
\newcommand\prd{{Phys.~Rev.~D}}%
\newcommand\apjl{{ApJ}}%
\newcommand\apj{{ApJ}}%
\newcommand{\pasp}[2]{{PASP}, {\bf #1}, #2}
\newcommand{\mnras}[2]{{MNRAS}, {\bf #1}, #2}
\newcommand\nat{{Nat}}%
\newcommand\apjs{{ApJS}}%
\newcommand\physrep{{Phys.~Rep.}}%
\newcommand\aap{{A\&A}}%


\title{New Limits on Gravitational Radiation using Pulsars}

\author{ A. N. Lommen } 
\institute{Astronomy Department \& Radio Astronomy Laboratory,
University of California, Berkeley, CA 94720-3411, USA\\
currently at Sterrenkundig Instituut, University van Amsterdam,
Kruislaan 403, 1098 SJ Amsterdam, the Netherlands\\
email: alommen@astro.uva.nl
}
\maketitle

\begin{abstract}
We calculate a new gravitational wave background limit using
timing residuals from PSRs J1713+0747, B1855+09, and B1937+21.  The
new limit is based on 17 years of continuous data pieced
together from 3 different observing projects: 2 at the
Arecibo Observatory and 1 at the 140ft Green Bank Telescope.
This project represents the earliest results from the
`Pulsar Timing Array' which will soon be able
detect the stochastic background from early massive black
hole mergers.
\end{abstract}

\setcounter{footnote}{0}
\setcounter{figure}{0}

\section{Introduction}

There are two categories of sources that could produce a detectable
level of stochastic gravitational radiation from the distant Universe.
One is of cosmological origin and the other involves galaxy evolution.
First,
low-frequency ($10^{-20{\rm~to~}0}$ Hz) relic gravitational radiation
may be generated during inflation, according to
string theory models of the early Universe,
via both early evolution of the extra dimensions and decay of
cosmic strings.
Cosmic strings, while now disfavored as the driver behind
structure formation in the Universe, still arise in many grand
unified theories of particle physics
after the epoch of inflation 
(Caldwell, Kamionkowski, \& Wadley 1998; Hogan 2000; see also Maggiore
2000).
Second, the coalescence of massive black holes (MBHs) during
galaxy evolution could also produce a detectable level of Gravitation
Waves (GWs).  (See Lommen \& Backer 2001 for more on this.)
Three cosmology experiments are being conducted or planned
by a number of investigators around the globe to either detect or
place new limits
on the stochastic background of
gravitational radiation:
polarization of the Cosmic Microwave Background Radiation (pCMBR),
Pulsar Timing Array (PTA),
and Laser Interferometer Space Array (LISA).

The PTA consists of a regular observing schedule on a handful of
the millisecond pulsars that are most accurate as clocks.  So far,
we have used the Arecibo Telescope in conjunction with the Green Bank
140ft telescope, but we plan on including the new Green Bank 100-m
telescope that has just been commissioned and a collection
of other European and Australian telescopes to maximize the observation
time and the range of observations available.

Pulsar timing residuals are sensitive to gravitational radiation
of wavelengths of about 1 yr, making the PTA complementary
to LISA which will measure much shorter wavelengths, and
the pCMBR which will measure much longer wavelengths.

Using timing residuals from Arecibo Observatory observations
of PSRs B1937+21 and B1855+09 over 8 yr,
Kaspi,Taylor, \& Ryba (1994, hereafter KTR94)
placed a limit of
$$
\Omega _g h^2 < 6 \times 10^{-8}   {\rm~~(95\% confidence)}
$$
where $\Omega _g$ is the fractional energy density in
gravitational waves per logarithmic frequency interval and
the Hubble constant H$_o = 100h$~ km s$^{-1}$ Mpc$^{-1}$.
We have connected the KTR94 data to 10 years of
Green Bank 140ft telescope data and 3 years of Arecibo Observatory data.
The overlapping data sets span a total of nearly 17 years, doubling
the baseline of Kaspi et~al's results.
The sensitivity of the PTA is proportional to $1/{(time span)^4}$,
and therefore
we expect an increase in sensitivity of about 16 over the KTR94 results.

First, in \S\ref{sec:observations} we describe the 3 different
sets of observations that have gone into calculating the limits
presented in this article.  
In \S\ref{sec:connection}
we connect the three data sets.
In \S\ref{sec:gr} we show the calculation of the new gravitational
wave limit.  \S\ref{sec:fractional_stability} compares the precision
of pulsar clocks to those of atomic time standards.
\S\ref{sec:timingnoise} is an update of the relationship
between $\dot{P}$ and timing noise presented by
Arzoumanian et~al. (1994).
In \S\ref{sec:planet} we discuss the possibility that a
planet orbits pulsar PSR B1937+21.
Finally in \S\ref{sec:conclusion} we present our
conclusions.

\section{Observations}
\label{sec:observations}

At the NRAO\footnote{The National Radio Astronomy Observatory (NRAO)
is operated by Associated Universities, Inc., under contract
with the National Science Foundation.} 140-foot (42.7 m) telescope
we observed PSR J1713+0747, PSR B1855+09, and
PSR B1937+21 in addition to others 4-6 times per year at
radio frequencies near 800
and 1400 MHz from 1989 October to 1999 July.
For details on the analysis of
these data see the report by
Backer et~al. (1993) on results from the first half of the data
set.
Observations of
PSR J1713+0747 started after its discovery
in 1994 (Camilo, Thorsett, \& Kulkarni, 1994).

We have been conducting monthly observations at 0.43 GHz,
1.4 GHz and 2.4 GHz of an array of
MSPs using the NAIC\footnote{The National Astronomy and Ionosphere Center
Arecibo Observatory is operated by Cornell University
under contract with the National Science Foundation.}
Arecibo Observatory 300 m telescope
since December 1997. These data are used
to make precision arrival time measurements for a variety
of astrophysical goals. We used the Arecibo-Berkeley
Pulsar Processor (ABPP), which is a multi-channel, coherent dispersion
removal
processor\footnote{`coherent' means that the dispersion is removed in
the voltage domain prior to power detection.}
with 112-MHz total bandwidth capability (For a detailed technical
description of the hardware see Backer et~al.(1997.)
For PSRs J1713+0747 and B1855+09
we use 56-MHz bandwidth for observations at 1.4 GHz, and 112-MHz for
2.4 GHz, while for PSR B1937+21 we use 45-MHz bandwidth at 1.4
GHz and 56-MHz at 2.4 GHz.

Calibrated total intensity profiles were formed from signals with
orthogonal circular polarization.
The profiles were then cross correlated with a template to
measure times of arrival (TOAs) relative to the observatory atomic clock.
Small errors in the observatory UTC clock,
of order 1 $\mu$s, were corrected based on
comparison of local time to transmissions from the Global
Positioning System of satellites (GPS).
The templates used for cross correlations were constructed by
fitting a set of Gaussian components 
to long-term averages of observations, using
the software described in Kramer et~al. (1994) and Kramer (1994). 
This model fitting scheme is described extensively in Lommen (2001).
We use the model to generate
noise-free templates with a specific common fiducial point
at all frequencies.  
Lommen (2001) shows that the templates generated
in this manner have at most 2 $\mu$s of error in the fiducial point
between 1400 and 2380 MHz.   

In addition to the two data sets mentioned above, we also have
incorporated the archival Arecibo data from KTR94 on PSR B1855+09
and B1937+21 into our analysis.  See the original paper for details
on these data.

\section{Analysis}
\label{sec:analysis}

As we just mentioned, TOAs are calculated
calculated via cross-correlation with a template.
Before these TOAs can be analyzed 
several corrections are required:
$$
TOA^\prime=TOA + \Delta_{clock} + \Delta_{DM} + \Delta_{backend}
$$
$\Delta_{clock}$ is comprised of a number of clock corrections
which in our case transform a TOA which is referenced against
an observatory maser, to one which is referenced against UTC.
$\Delta_{DM}$ causes TOA$^\prime$ to be the time that the pulse
would have arrived had it been at infinite frequency (lower
frequencies are delayed more by the ISM).
$\Delta_{backend}$ corrects for any delays that are dependent
on the observing backend used.  In our case we have a digital latency
which is dependent on channel bandwidth
and a mid-scan correction which is a result of our on-line folding.
After all of these corrections,
TOA$^\prime$ is independent of observatory clock, frequency,
and observing backend.  TOA$^\prime$ marks the space-time event
of the arrival of a defined fiducial point of the pulse at the
telescope on a defined time scale.

\section{Connection of Data Sets
}
\label{sec:connection}

In the TEMPO analysis arrival time of
pulses at the observatory is referred to that
at the barycenter of the solar system.  
The solar time corresponding to the rotation of the earth is
known as UT1.  UT1 is tabulated by measuring the difference
between UT1 and UTC.  UTC is a solar time scale that
runs at the rate of TAI, international
atomic time.  BIPM tables of (UT1-UTC) are used to orient the
observatory in inertial space and make the translation from observatory to
earth-center.  (UT1-UTC)
is between 7 and 32 seconds over the course of our observations, and
is known to 0.1 ms at all times.
This translates to an uncertainty in TOA of only $\sim$ 0.1 ns.
To calculate the earth's position TEMPO uses standard
JPL ephemerides DE405.1950.2050.
We know the geodetic position of
each observatory to within meters.
The delays that remain, that
are most difficult to account for, are the delays between the arrival
of the pulse at the fiducial point of the antenna as defined by VLBI
coordinates and the completion of its transmission through
the `back-end' of the receiving system.  These are typically in
the range of 100-3500 ns.  At Arecibo, for example, we know that the
travel time for a signal from the control room to the Gregorian
receivers and back is approximately
7 $\mu$s \footnote{Mike Nolan, private communication,
2001 Aug 19} largely resulting from the $\sim$1200 ft of fiber-optic
cabling through which the signal must travel from the platform
to the control room.  This would give us a $\sim 3.5~\mu s$ delay
in the TOAs. Typically, we ignore these
delays because we take data from a single observatory, with a single
operating system, and any constant delay such as this is seemingly
irrelevant.

However, these delays become important for 
connecting multi-observatory data sets. 
If all such delays at all observatories engaged in pulsar timing
were known, it would be possible to have a ``Universal Pulsar Timing
Array (UPTA)'' in which all TOAs from all pulsar data from all over
the world could be combined a priori in single meaningful data base.  This
would necessitate the need for uniform template fiducial point
processing.  For example, if the world's pulsar astronomers could piece
together 17-year data sets on 10 different MSPs with comparable
precision, the sensitivity of the UPTA would go down by a factor
of 1/$\sqrt{10}$.

In our comparatively small experiment, we attempt this, but
can only account for everything but the final 15-60 $\mu$s,
and so in the end we end up
doing a bootstrap
connection.  We have 3 overlapping data sets.
For each pulsar, the average separation between two overlapping sections
is used
as the offset between the two data sets.  These offsets, which
will be included
in the ITOA format that we will make publicly available,
are shown for each pulsar and for each adjacent data set in
Table \ref{tab:offsets}.  In the case of the Green Bank and ABPP
data sets, the TOAs were generated using the same series of Gaussians
with the same fiducial point, so all template discrepancies are
removed to within 2 $\mu$s (see Lommen 2001 
for more on this subject).
2 $\mu$s in fact, represents the maximum error in going between
1400 and 2380 MHz.  The data we have connected 
is centered from 1330 MHz to 1420 MHz, and the error roughly
scales with the size of the bandwidth over which one must construct
templates (see Lommen 2001).  Therefore the 
maximum error in connecting our range of data, due to our
template cross-correlation is roughly a tenth
of 2 $\mu$s, or 0.2 $\mu$s.
In the case of the offset between the KTR94 data sets
and the Green Bank data sets, we added, a priori, an amount of
offset corresponding to the difference between our fiducial point
and the fiducial point of KTR94, who used the peak of the template.
These a priori numbers were a phase of 0.443259 for B1855+09 and
0.353460 for B1937+21.  The numbers shown in the table, are the
offsets in addition to these template offsets.
The sign of the offsets are meaningful, e.g. the numbers under
the column ``GB-ABPP" are the quantities acquired from subtracting
the average ABPP residual from the average Green Bank residual during
the epoch while they overlap.  
The offsets between data sets that we processed ourselves
(GB - ABPP) are all less than 8 $\mu$s.   

\begin{table}
\caption[Offsets in time between different observatories] {
\label{tab:offsets}
Offsets, in $\mu$s, between data sets.}
\begin{center}
\begin{tabular}{||l|c|c||} \tableline 
Pulsar & KTR94-GB\tablenotemarka & GB-ABPP\tablenotemarkb \\ \tableline
J1713+0747 & \nodata & $-3.2 $ \\ \tableline
B1855+09 & $-56.9 $ & $ 1.4 $ \\ \tableline
B1937+21 & $-14.4 $ & $-0.12 $ \\ \tableline
\end{tabular}
\end{center}
\begin{list}{}{}
\tablenotetexta {GB=Green Bank data}
\tablenotetextb {ABPP=ABPP data (post-upgrade Arecibo)}
\end{list}
\end{table}

In order to perform a weighted fit across multiple data sets, it is
necessary to be very careful of the average weighting of one data
set relative to another.  Said another way, with a single set of
data, all taken with a single observing system and analyzed uniformly,
one need not worry about, for example, overestimating all the error
bars by a factor of 3.  However, when one is combining two such data
sets, if one has overestimated error bars, and a second has underestimated
error bars, the fit will tremendously favor the latter.   In order to
account for this problem, we normalized the error bars in each data
set by the total variance per day in those data, i.e.
$$
\sigma_{\rm normalized} = \sigma_{\rm original}\times
\frac{\sqrt{\sum_{i}{1/{\rm err}_i^2}}}
{L}
$$
where $L$ is the length of the data set, in days, and
$\sigma_{original}$ are the error bars that were published
by KTR or calculated by us, using the method created by
Christophe Lange (private communication)
which was adapted from Downs \& Reichley (1983). 

We used the standard TEMPO
package\footnote{http://pulsar.princeton.edu/tempo}
to perform weighted fits to
$\alpha$, $\delta$, $\mu_\alpha$, $\mu_\delta$, $P$, and $\dot{P}$,
in each of the 3 pulsars.
Additionally
in J1713+0747 and B1855+09 we fit for the 5 Keplerian parameters and
the Shapiro delay parameters, m$_2$ and
$\sin{i}$ (See Stairs, 1998 for an excellent summary
of the process of fitting for the Shapiro delay).

\begin{figure}
\centerline{\psfig{file=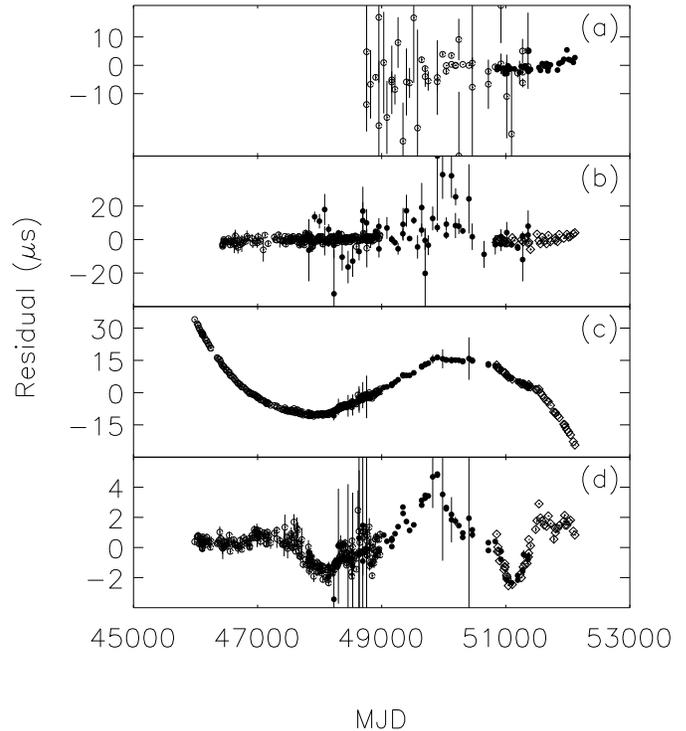,width=8.8cm}}
\caption[Timing Residuals PSRs J1713+0747, B1855+09, B1937+21]{
\label{fig:master_residual}
(a) PSR J1713+0747, (b) PSR
B1855+09, (c) PSR B1937+21 and (d) PSR B1937+21 fit for $\ddot{P}$.
The open circles are KTR94 data, the filled circles are Green Bank data, 
and the open diamonds are ABPP data.  
}
\end{figure}

We show the resulting residuals in Figure \ref{fig:master_residual}.  In
(a),
(b), and (c), we show J1713+0747, B1855+09, and B1937+21, respectively,
fitted for the parameters not marked with footnote ``c" in the table.
Additionally, in panel (d), we show B1937+21 fitted for
$\ddot{P}$($\ddot{\nu}$).

In the next three sections, we discuss various implications of
the residuals resulting from these fits.  
First, we assume the residuals shown in Figure \ref{fig:master_residual}
are the result of GWs and use them
to place new limits on the GWB.  Second, we assume the
same residuals are the result of timing noise intrinsic to
the pulsars, and determine whether these pulsars are more or
less noisy than expected based on the characteristics of the
population at large.
Finally, we consider the possibility that the residuals in
B1937+21 are the result of a planet orbiting that body.

\section{Limit on Gravitational Radiation}
\label{sec:gr}

These data, representing a continuous 17-y set,
provide a new limit on the level
of background gravitational radiation present throughout the galaxy,
and ostensibly the universe.  In short, we assume that all
the observed fluctuations in the timing residuals are due to
the stochastic background of gravitational radiation.  This
yields the maximum possible gravitational wave spectrum present.

To determine the limit they place, we first need to measure the
spectrum of fluctuations that we observe in the timing residuals.
The process of computing an observed
noise spectrum from irregularly sampled data has been given much
consideration by previous authors 
(Groth 1975; Deeter \& Boynton
1982; Deeter 1984; Cordes \& Downs 1985; KTR94).
These authors are particularly concerned with ``red'' spectrum
where imperfect sampling leads to coupling of spectral estimates
at low frequency.
Deeter (1984) compares the various methods and determines
that the method of fitting the data to orthonormal polynomials
produces the most meaningful results.  We therefore use this
method, as did both
Stinebring et~al. (1990) and KTR94 in previous generations
of this experiment.

Stinebring et~al. (1990) and KTR94 start with a calculation
which shows that the energy density of background gravitational
radiation per frequency octave would have
a frequency dependency $\propto \nu^{-5}$ and they go about putting limits
on
that assumed spectrum.  We do the same.  However, Rajagopal \& Romani (1995),
Phinney (2001), and Jaffe \& Backer (2002)
recently show that the energy density per frequency octave
would be proportional to
$\nu^{-4.3}$ rather than $\nu^{-5}$.  Either way, this is steep
compared to the intrinsic spectrum observed in young pulsars,
and would not change our
results significantly.

A series of polynomials, $p_i(t)$ where $i$ is of the set ${0,~1,~2,~3}$
of corresponding order, is generated
orthogonally over the sampling of the timing residuals.  We
started with the following set:
$$
p_{j=0,1,2,3}(t)=\sum_{i=0}^j{t^i}
$$
The
process of orthogonalization is done via standard Gram Schmidt reduction.
A linear combination of the $p_i$'s is found which minimizes
the quantity
$$
\chi^2=\frac{1}{\sigma^2}\sum_{n=1}^N\left(r(t_n) -
\sum_{j=0}^3C_jp_j(t_n)\right)^2
$$
where $r(t_n)$ is the measured residual at time $t_n$.
The first three $C_j$'s are covariant with the fitting of the
phase, period, and period derivative in the pulsar model.  $C_3$,
however, is a measure of the amplitude of the spectral density
of the variance in the timing residuals at a frequency
corresponding to 1/T where T is the length of the data set.
$S_m=<C_3^2>$ is the corresponding estimate of the power(variance)
contained near frequency 1/T.
In order to sample the spectrum of $S_m$ over
multiple frequencies, we divide the data into smaller and
smaller subsets in time, by factors of 2.  We refer to the divisor
as the frequency index.  A frequency index of 2 implies the
data set was divided in half, and that therefore we are measuring
frequencies of 2/T, or periods of about 8.5 years.

\begin{figure}
\centerline{\psfig{file=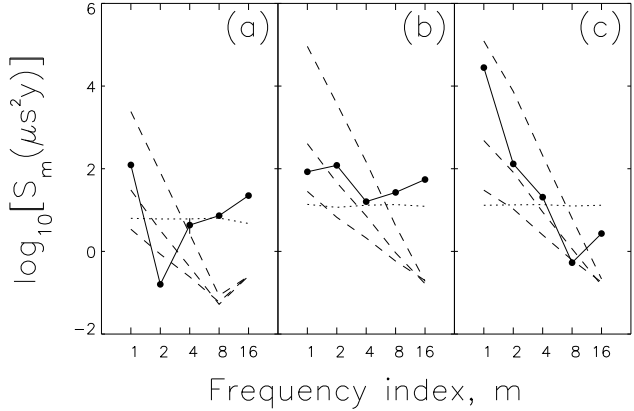,width=8.8cm,clip=}}
\caption[$\log{S_m}$ vs frequency index]{
\label{fig:kaspifig2}
$\log{S_m}$ vs frequency index.  Solid line shows measured
values for each pulsar.  Dotted lines show simulated data for
spectral index 0.  Dashed lines show simulated data for
spectral indices 2, 3, and 5.  (a) PSR J1713+0747, (b) PSR
B1855+09, and (c) PSR B1937+21.  Frequencies corresponding to
frequency index are m/L where L is the length of the data
set: (a) 9.2 yr, (b) 15.6 yr, and (c) 16.8 yr.
}
\end{figure}

The results of the spectral measurements are shown in the
solid line
Figure \ref{fig:kaspifig2} for PSRs J1713+0747, B1855+09,
and B1937+21, using the fits corresponding to parts a, b, and c
of Figure \ref{fig:master_residual}, i.e. fits through $\dot{\nu}$ only.   
Though this method works well for irregularly
sampled data, the value of the $S_m$s are still
dependent on sampling.  In order to gauge the extent
of this dependence we have simulated data sets of various
spectral indices, with the same sampling as the actual data.
Randomly distributed Gaussian noise was transformed into
the Fourier domain, multiplied by a function with spectral
index 0,2,3, or 5, normalized to 1~$\mu$s$^2$~y at a frequency
of 1 y$^{-1}$, and transformed back to a time series.  Once in
the time domain, the data were sampled identically to
the actual data. 10,000 such
realizations were created for each of the 4 spectral indices.
The power spectra of each was measured using the method
of orthonormal polynomials described above.  The results of
this analysis are shown in dotted lines in Figure
\ref{fig:kaspifig2} for each of the 4 spectral indices.

The average value of the spectral estimator, $S_m$, with spectral index
5 represents the estimate of $S_m$ in the presence of a gravitational
radiation, $<S_m(\Omega_gh^2)>$, where $\Omega_g$ is the energy
density in GWs, $\rho$, expressed as a fraction of the closure 
density, $\rho_c$ of the universe:  
$$
\Omega_g= \frac{\rho}{\rho_c} = \frac{8\pi G\rho}{3H_o^2} 
$$
where $H_o$ is the Hubble constant.
We generally express $\Omega_g$ in terms of $h$, i.e. $\Omega_gh^2$
where $H_o = 100~h~{\rm km~s}^{-1}~{\rm Mpc}^{-1}$.
We need to translate the normalizing
amplitude, (1~$\mu$s$^2$~y at a frequency of y$^{-1}$) into a particular
value of $\Omega_gh^2$.  We use
the relationship derived from the definition of $\Omega_g$
 (from Stinebring et~al. 1990)
$$
P_g(f)=\frac{H_0^2}{8\pi^4}\Omega_gf^{-5}.
$$
where $P_g(f)$ is spectral density, i.e. the same quantity
we are attempting to measure with $S_m$.  
This means that values of $<S_m(\Omega_gh^2)>$ from data
with 1~$\mu$s~y at a frequency of y$^{-1}$
need to be divided by 752 in order to
represent the spectral density when $\Omega_gh^2 = 10^{-7}$ which
is the value we represent in Table \ref{tab:Smg}.

Figure \ref{fig:kaspifig2} shows that the measured spectra
of both J1713+0747 and B1855+09 are both quite flat compared
to the spectral index 5 model.  
We would expect J1713+0747 to be somewhat flatter than the others
simply by virtue of the plot representing a higher frequency region of
the spectrum.  PSR B1937+21 demonstrates a steep spectrum, comparable
to the spectral index 5 model spectrum, but this may be due to
other sources as we will discuss later (e.g. see \S \ref{sec:planet}).

In order to calculate the expected influence of gravitational
waves on the spectral measurements, $S_m$, we needed to know
the contribution from purely white noise.  To this end, we simulated
residuals that were only influenced by our known
quantities of instrumental noise.  We created
white noise at the same sampling as that of the three pulsars
we are considering.
The white data from each observatory were normally distributed
random numbers normalized such that
their RMS matched the median of instrumental noise from that
pulsar at that observatory.  The value used for instrumental
noise was the standard deviation of the mean
of the timing residuals about
the mean for a particular day, i.e.
$$
stdev = \frac{\sqrt{\sum_i(r_i - u)^2}}{N}
$$
where $r_i$ is a single residual of which there are $N$ averaged
to form a daily average and $u$ is the mean of the residuals on
that day.
We created 10,000 random
realizations for each of the 3 pulsars, and measured their
spectrum using the same technique shown above.  The average
value of these data we call $<S_m>_w$.

The values of $S_m$, $<S_m>_w$, and $<S_m>_g$ for each value
of the frequency index, m, are given in Table \ref{tab:Smg}.

\begin{table}
\caption {
\label{tab:Smg}
Observed and computed spectral densities.}
\begin{center}
\begin{tabular}{||l|c|c|c|c||} \tableline
Pulsar & m & $S_m$ & $<S_m>_w$ & $<S_m>_g$ \\ \tableline
J1713+0747 & 1 & 348  &538 & 3.21 \\ \tableline
    & 2 & 6.20 &435 & 0.108 \\ \tableline
    & 4 & 22.4  &410 & $3.52 \times 10^{-3}$ \\ \tableline
    & 8 & 4.17 &420 & $1.14 \times 10^{-4}$ \\ \tableline
    & 16 & 111 &278 & $3.34 \times 10^{-4}$ \\ \tableline
B1855+09   & 1 & 84.8 &137 & 121 \\ \tableline
    & 2 & 120  &182 & 5.01 \\ \tableline
    & 4 & 15.6 &210 & 0.195 \\ \tableline
    & 8 & 26.6 &204 & $5.57 \times 10^{-3}$ \\ \tableline
    & 16 & 54.7 &155 & $2.51 \times 10^{-4}$ \\ \tableline
B1937+21   & 1 & 29379 &0.025 & 164 \\ \tableline
    & 2 & 203 &0.028 & 9.87 \\ \tableline
    & 4 & 11.8 &0.035 & 0.277 \\ \tableline
    & 8 & 7.93 &0.035 & $8.39 \times 10^{-3}$  \\ \tableline
    & 16 & 3.43 &0.034 & $2.87 \times 10^{-4}$ \\ \tableline
\end{tabular}
\end{center}
\begin{list}{}{}
\tablenotetexta {GB=Green Bank data}
\tablenotetextb {ABPP=ABPP data (post-upgrade Arecibo)}
\end{list}
\end{table}

In order to attempt to rigorously
calculate an upper limit to background gravitational radiation
present in our residuals, we duplicate the analysis of
Thorsett \& Dewey (1996, hereafter TD96).   
There is significant controversy around
this method, which we address following the results.
TD96 create a statistic,
$Stat_1=mS_m/[<S_m>_w + <S_m(\Omega_gh^2)>_g]$,
which is normally distributed with $m$ degrees of freedom.  The
statistic is essentially the measured value of the spectral estimator
divided by the predicted value, based on Monte Carlo simulations.

\begin{figure}
\centerline{\psfig{file=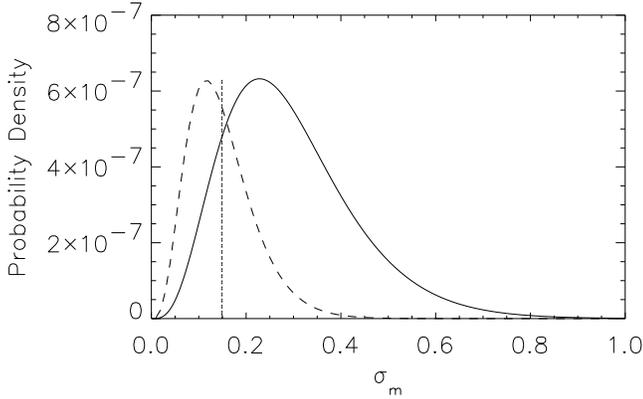,width=8.8cm,clip=}}
\caption{
\label{fig:thorsett_example}
An example of two different probability density distributions.
5\% of the area under the solid curve is to the left of the dotted
line.
}
\end{figure}

We use a Neyman-Pearson test to determine whether the distribution
of $Stat_1$, for various values of $\Omega_gh^2$ is significantly
different than the corresponding statistic in the absence of
gravitational radiation.  The validity of the Neyman-Pearson test
is complicated, but it has a simple construction.  Consider a particular
measurable, A, with probability distribution S1.  Correspondingly,
consider another measurable, B, with probability distribution S2.
The Neyman-Pearson test essentially tests the extent to which
the distributions overlap.  If they have considerable overlap, they
are indistinguishable, if not, they aren't.  The quantitative test
of `considerable' involves computing the line which delineates
5\% of the area under the curve, representing
the least likely outcomes of measuring A.  Consider Figure
\ref{fig:thorsett_example} which shows two probability distributions.
The area to the left of the dotted line is 5\% of the total area
of the solid curve.  The mean value of the dotted curve, falls
just to the left of the dotted line.  Thus, we can say that these
two distributions are different from each other at the 95\%
confidence level.

One of the subtleties of the Neyman-Pearson test is that one
of the two hypothesis is considered the `null' hypothesis, and
this is the hypothesis one accepts unless the data compels one
to choose otherwise.  Thus, the Neyman-Pearson test, by
conjecture has a preferred answer, i.e. whatever you choose as
the null hypothesis.  Our null hypothesis will be that there
is no gravitational radiation in the universe.  This is of course
incorrect.  This is the first issue McHugh et~al. (1996) have with this
method.  We are inclined to agree that this is not a good construction.

TD96 used a likelihood ratio, which is an unnecessarily complicated
way to construct the Neyman-Pearson test.  We merely need to compare
the two distributions that are shown in the numerator and the
denominator of the ratio TD96 use.
$$
S1 = \Pi_{m=1,2,4,8} \chi^2_m\left(\frac{mS_m}{<S_m>_w +
 <S_m(\Omega_gh^2)>_g}\right)
$$
$$
S0 = \Pi_{m=1,2,4,8}\chi^2_m\left(\frac{mS_m}{<S_m>_w}\right).
$$

\begin{figure}
\centerline{\psfig{file=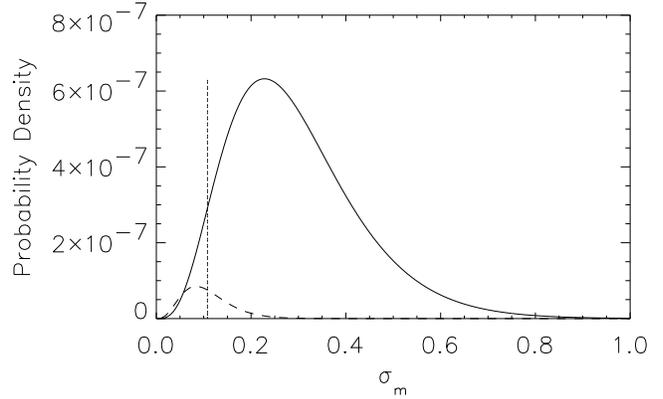,width=8.8cm,clip=}}
\caption{
\label{fig:thorsett}
The solid line shows the probability density of the null
statistic, S0, while the dashed curve shows the probability
density of the S1 statistic at the 95\% confidence value.
The dotted line is the same as in Figure \ref{fig:thorsett_example}
}
\end{figure}

S0 represents the null statistic, and is shown in the solid line
of Figure \ref{fig:thorsett}.  S1 represents the statistic when
a certain amount of gravitational radiation $\Omega_gh^2$ is
present.  We have plotted S1 in Figure \ref{fig:thorsett} when
its value is equal to the 5\% limit that we calculate, namely,
$\Omega_gh^2 \le 2.8 \times 10^{-6}$ at a 95\% confidence level.

TD96 calculated 
$\Omega_gh^2 \le 1.0 \times 10^{-8}$ at a 95\% confidence level 
using only the KTR94 B1855+09 data.
Using the technique of TD96
our expanded set of B1855+09 data actually
place a weaker limit than the KTR94
data alone, because of the large scatter in the Green Bank data points.
The technique separates the data into smaller and smaller sections
in order to measure shorter frequencies.  The average of these shorter
frequency sections is therefore falsely enlarged in our calculations.
Using only the $m=1$ term of our data we have:
$\Omega_gh^2 \le 9.8 \times 10^{-6}$ at a 90\% confidence level.

McHugh et~al. (1996) rejected the analysis of TD96 because
of their use of hypothesis testing rather than parameter estimation.
They completed a full Bayesian
analysis of the same data which resulted in a much less restrictive
limit, $\Omega_g h^2 < 9.3 \times 10^{-8}$, than the 
analysis of TD96 using the same data.

These opaque calculations cover up the essential nature of the
results.  After KTR94 we can make a much more simple estimate of
the limit these data place on the GWB.  We use KTR94 equation 6 which
relates the energy density, $\rho$, in gm cm$^{-3}$ of a gravitational wave to
its affect on TOA in $\mu$s, $A$, and a frequency, f, in Hz.
$$
\rho=\left(\frac{243\pi^3}{416G}\right)A^2f^4.
$$
The largest amplitude sinusoid that one could conceivably
fit to the B1855+09 data has $A=3~\mu$s and $f=1/17~yr = 1.9 \times 10^{-9}$
Hz.
This gives $\rho = 3.2\times10^{-38}$ gm cm$^{-3}$, or
using
$$
\frac{\rho}{\rho_c} = \frac{8\pi G \rho}{3H_o^2} =2 \times 10^{-9}h^{-2},
$$
which is more than an order of
magnitude smaller than the limit found by KTR94.  We assume
$H_o = 100~h$~km~s$^{-1}$~Mpc$^{-1}$.

As explained in Lommen \& Backer (2001),
a gravity wave propagating through the galaxy would have an effect
both on the emitting site (the pulsar) and the receiving site (the Earth).
Neglecting geometrical considerations for a moment, this could
imply that the limit we have just derived will on average be a factor
of 2 too high, i.e., that the response that we see in the pulsar
timing, since it results from effects at both sites, is essentially
doubled.   The factor is not so simple, but depends on (a) the
direction of propagation and polarization of the impinging gravitational
wave, and (b) the geometrical relationship of the pulsars to the Earth.
For a stochastic background of gravitational waves, however, we can
calculate the expected effect.
If the GW at the emission and reception sites constructively
interfere then $C_3$ will increase by a factor of 2, and $S_m$ will
increase by a factor of 4.  If emission and reception destructively
interfere then $S_m$ will be close to 0 (it will certainly be
non-zero due to irregular sampling). 
The expectation value, i.e. the ensemble average
over the 4$\pi$ directions of propagation, and over the wavelengths
in question is a factor of 2, so the limit is actually half what 
we, or any of these other groups, measure.  
This of course leaves aside the
issue of the variance in the GWB.  The background value we measure using
these data place a limit on the GWB in a specific place and epoch:
in our galaxy and during the 2 decades in which
this experiment took place.  
Jaffe \& Backer (2002) are working on a 
realistic version of the GWB which takes these specifics into account. 

\begin{figure}
\centerline{\psfig{file=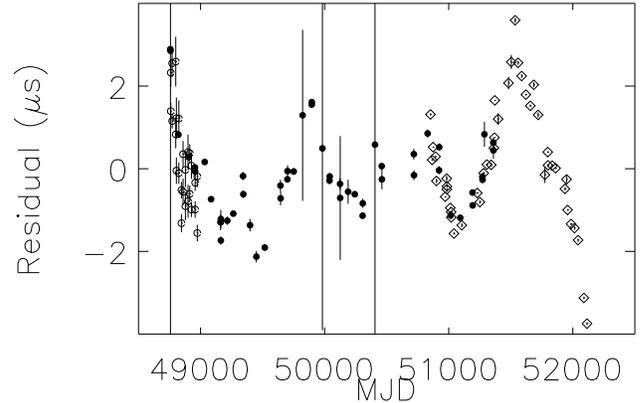,width=8.8cm,clip=}}
\caption{
\label{fig:1937_1713range}
Residuals from B1937+21 with no $\ddot{\nu}$ removed, fitted only
to the range of the J1713+0747 data.
}
\end{figure}

We have assumed up until now, that PSR B1855+09 produces the best
limit on the GWB since it demonstrates the most
stable timing and a long baseline.  However, what if there was
a geometrical situation in which B1937+21 was experiencing a large
perturbation due to GWs while a null occurred at PSRs
J1713+0747 and B1855+09?
We first need to consider that the limited range of data on J1713+0747 was
preventing our detection of any cubic therein.  In Figure
\ref{fig:1937_1713range} we performed the same fit that we
did for part (c) of Figure \ref{fig:master_residual} but using
only the limited range of the J1713+0747 data.  The obvious
cubic is gone but there is a `sawtooth' present at the level
of 2~$\mu$s that we do not see in the J1713+0747, so we conclude
that we are seeing no comparable gravitational wave in J1713+0747.

\begin{figure}
\centerline{\psfig{file=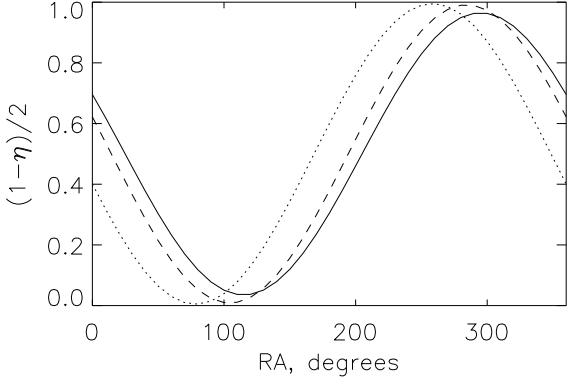,width=8.8cm,clip=}}
\caption[$(1-\gamma)/2$ vs RA for DEC=0 for 3 pulsars]{
\label{fig:find_nulls}
$(1-\gamma)/2$ vs RA for DEC=0 for each of the 3
pulsars.  B1937+21 is shown by the solid line.  J1713+0747 is the
dotted line, and B1855+09 is the dashed line.
}
\end{figure}

To attempt to find a geometrical situation which produces
an approximate null in J1713+0747 and B1855+09
we inspected the range of possible geometrical multiplicative
factors, $(1-\gamma)/2$ over the 4$\pi$ sphere of possible gravitational
wave directions for each of the three pulsars.  See Lommen \& Backer (2001) for
an explanation of this factor.  Figure \ref{fig:find_nulls} shows
the factor for 2$\pi$ values of RA while DEC=0\de for each of the 3
pulsars.  B1937+21 is shown by the solid line.  J1713+0747 is the
dotted line, and B1855+09 is the dashed line.  The three pulsars are
close enough together in the sky that $(1-\gamma)/2$ is highly
correlated over the sphere.  We chose the DEC=0 line to show because
it essentially produces the most favorable spot on the sphere for
a B1937+21 gravity wave enhancement with a near-null in J1713+0747
and B1855+09.  At DEC=80\de J1713+0747 has $(1-\gamma )/2=0.005$, B1855+09
has $(1-\gamma)/2=0.05$, while B1937+21 has $(1-\gamma)/2=0.12$.
That gives B1937+21 a 2.4$\times$ enhancement over B1855+09 and
a 24$\times$ enhancement over J1713+0747.  The relative strengths
of J1713+0747 and B1855+09 can be exchanged somewhat as you can
see in Figure \ref{fig:find_nulls}.
We conclude that it would be very difficult to arrange for the
`deviant' residuals we see in B1937+21 to be produced by
gravitational radiation without producing a similar, but
slightly smaller, $\sim 10~\mu$s effect, in either J1713+0747 or B1855+09.
In addition, a gravitational wave disturbance, which, when reduced
by a factor $(1-\gamma)/2=0.12$ still yields a 30~$\mu$s residual
deviation is unrealizable using any known gravitational
wave producer in the universe (see Lommen \& Backer 2001).

Finally, we consider whether emission and reception site disturbances
could be destructively interfering in the case of 
B1855+09 
and constructively interfering in the case of PSR B1937+21.
The stochastic GWB can be thought of as ``crinkled'' space-time.
We approximate the crinkling as a sine-wave of equal amplitude 
but arbitrary phase near each of the 2 pulsars and also near
the earth.
We begin by computing the enhancement of the amplitude of a sine
wave by adding two equal sine-waves together, with a variable
displacement, $\phi$, between the two.  This enhancement we call $E(\phi)$.
Thinking of B1937+21 and the earth
as emitter and receiver, we may, for example,
calculate the probability that $E(\phi)$
will be 5 or greater:  0.001.
To obtain the probability that B1937+21's perturbation will be
enhanced over B1855+09's perturbation by a factor of 25 (roughly
the ratio of their peak deviations)  or
greater we compute the quotient:
$$
Q(\Delta\phi)=\frac{E(\phi)}{E(\phi - \Delta\phi)}
$$
for $\Delta\phi=[0,2\pi]$.
The probability that $Q(\Delta\phi) > 25$, 0.016, is the probability that
the disturbance in B1937+21 is 25 times that in B1855+09.
Thus, the residuals
we observe in B1855+09 render it highly unlikely that
the large amplitude quasi-sinusoid observed in B1937+21 is
the result of a GW.  Consideration of PSR J1713+0747 would 
suppress the probability further but not by an equal factor.

\section{Fractional Stability of Terrestrial Atomic Time Standards vs. Fractional Stability of Neutron Star Rotations}
\label{sec:fractional_stability}

\begin{figure}
\centerline{\psfig{file=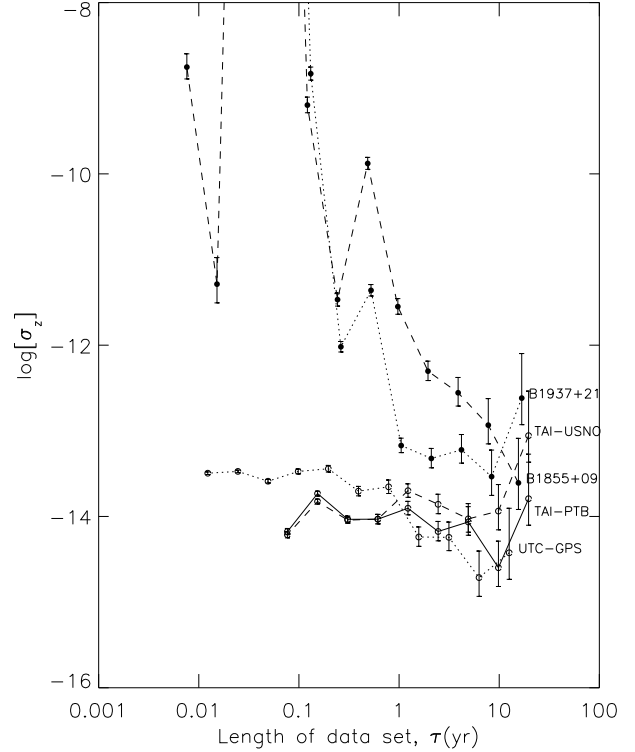,width=8.8cm,clip=}}
\caption{Fractional Stability of 3 terrestrial timescales, and 
2 MSPs.
\label{fig:sigmaz}
}
\end{figure}

At the baselines considered in this paper, namely longer than
10 years, the fractional stability of neutron star rotations rivals that
of terrestrial atomic time standards.  To make the comparison quantitative
we used the statistic $\sigma_z$ proposed by Matsakis, Taylor, \& Eubanks 
(1997) 
for describing pulsar and clock stabilities.  The recipe for
computing $\sigma_z$ given in Matsakis, Taylor, \& Eubanks (1997)
is very complete.
We only give a bare outline of the computation process here.
We divide the timing residual or clock comparison, $X(t)$, 
into smaller and smaller subsections of
time, starting with the full length, T, and going to T/2, T/4, T/8, etc.
To each subset we fit the function
$$
X(t) = c_o + c_1(t-t_0) + c_2(t-t_0)^2 + c_3(t-t_0)^3.
$$
$\sigma_z$ is related to the average of $c_3^2$ by the following
formula
$$
\sigma_z(\tau) = \frac{\tau^2}{2\sqrt{5}}<c_3^2>^{1/2}
$$
where the angular brackets denote the average, and $\tau$ is
the length of the subset of data.
Figure \ref{fig:sigmaz} shows the fraction stability, $\sigma_z$ 
vs. dataspan for [UTC-GPS], [TAI-PTB], [TAI-USNO], and [PSR-TAI]
for PSRs
B1855+09, B1937+21.  To compute the error we used the
estimate provided by Matsakis, Taylor, \& Eubanks (1997), which
underestimates the error in the case of data with significant
red noise, as is evident in the pulsar data.

Figure \ref{fig:sigmaz} shows that PSR B1855+09 beats
the [TAI-USNO] time scale at the longest time periods, and
is essentially equally as accurate as [TAI-PTB] at timescales
of 19 years.  Since the fractional stability of B1855+09 improves
steadily
with increased dataspan, we expect that as additional data on 
B1855+09 is acquired, this pulsar will easily beat [TAI-PTB] at $\sim$25
year timescales unless TAI and PTB improve or the pulsar becomes less stable. 
Note that TAI is significantly better than USNO alone even though
TAI is an aggregate that contains USNO, and similarly, for TT(BIPM) which
contains TAI.  It is also interesting to note that 
the PTA allows the possibility of ignoring the
earth clock altogether; one degree of freedom in the PTA can be used
to solve for time, i.e., PSR vs PSR.

\section{Update on Timing Noise as a Function of Period Derivative}
\label{sec:timingnoise}

Pulsar astronomers have long wondered whether the noise 
they see in timing residual plots, such as we show in Figure
\ref{fig:master_residual} is due to intrinsic instability in the
rotation of the pulsar, or is an error in the model, such
as a missing planet or unmodeled DM variations.
Arzoumanian et~al. (1994) published a compendium of
measurements of ``timing noise'' in a number of slow and millisecond
pulsars, and suggested that there is a relationship between
timing noise and $\dot{P}$, i.e., the larger the $\dot{P}$
the larger the timing noise.  This implies first that the noise
is intrinsic to the pulsar, and second that for
each pulsar there is a minimum RMS in the residuals that no
amount of accounting for systematics will lessen.
This has obvious implications for the PTA, and our effort to
achieve sub-$\mu$s RMSs on a collection of MSPs.
The data presented here along with recently published results on
MSPs allow us to update the picture of timing noise in MSPs as
a function of period derivative.  This was most recently done by
Arzoumanian et~al. (1994) who quantified the noisiness of a pulsar
using the noise parameter, $\Delta_8$, defined by
$$
\Delta_8 = \log{\left(\frac{1}{6\nu}|\ddot{\nu}|t^3\right)},
$$
where $t=10^8$~s.  Arzoumanian et~al. (1994) quote the best fit
result to the scatter plot
$$
\Delta_8=6.6 + 0.6\log{\dot{P}}
$$
which brought forth the rather depressing notion that we would
never achieve better than $1~\mu s$ accuracy for the MSPs
owing to the fact that
$\dot{P}$ for the population is $\sim 10^{-20}$ or higher.

\begin{figure}
\centerline{\psfig{file=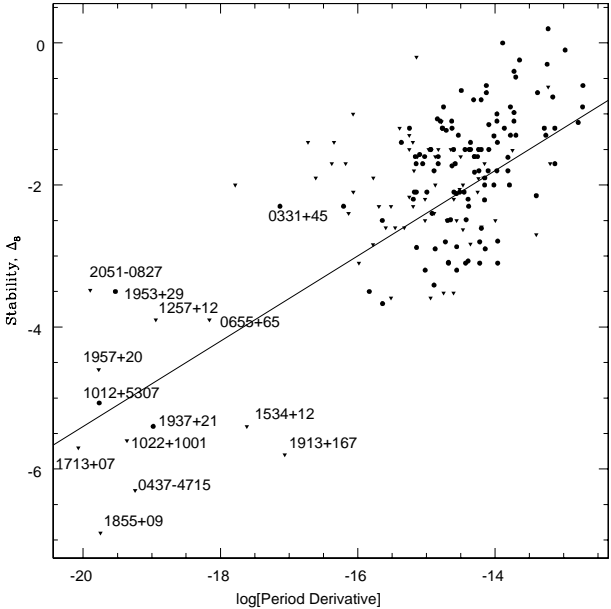,width=8.8cm,clip=}}
\caption{
\label{fig:zaven}
The timing noise parameter, $\Delta_8$, as a function of
$\log{\dot{P}}$.  Values used are courtesy of Zaven Arzoumanian,
except those shown in Table \ref{tab:zaven}, and 
1957+20 (Arzoumanian, Fruchter, \& Taylor, 1994),
B1534+12 (Stairs et~al, 1998).
}
\end{figure}

We are happy to report that some of the MSPs used to create the low
$\dot{P}$ end of the spectrum have been determined to
have lower $\Delta_8$'s than were presented in Arzoumanian et~al. (1994).
We present the new MSP data in Table \ref{tab:zaven} and
the corresponding plot in Figure \ref{fig:zaven}.

\begin{table*}
\caption[Timing Noise Parameter vs $\ddot{\nu}$] {
\label{tab:zaven}
Timing Noise Parameter vs $\ddot{\nu}$ for the MSPs for
which they are available.}
\begin{center}
\begin{tabular}{||l|c|c|c|c||} \tableline
Pulsar & $\dot{P}$ & $\ddot{\nu}$ (s$^{-3}$) & $\Delta_8$ & Source \\ \tableline
J0437+4715 & 5.72906(5)  &  $<5\times10^{-28}$ & $<$-6.3 & vs\tablenotemarka \\ \tableline
J1012+5307 & $1.7134(1)\times10^{-20}$ & $-9.8(2.1)\times10^{-27}$ & -5.1 & la\tablenotemarkb \\ \tableline
J1022+1001 & $4.341(4)\times10^{-20}$ & $<1\times10^{-27}$ & $<$-5.6 & kr\tablenotemarkc \\ \tableline
B1257+12 &  $11.4223(7) \times 10^{-20}$ & $<-1.35 \pm 0.04 \times 10^{-25}$
& $<$-3.9 & wz\tablenotemarkd \\ \tableline
J1713+0747 &  $8.54 \times 10^{-19}$ & $<-2.6 \pm 0.1 \times 10^{-27}$
\tablenotemarkc
& -5.6 & tw\tablenotemarke \\ \tableline
B1855+09 & $1.78 \times 10^{-18}$ & $1 \pm 6 \times 10^{-29}$ & $<-6.9$ &
tw\tablenotemarkf \\ \tableline
B1937+21 & $1.05 \times 10^{-19}$ & $1.515 \pm 0.001 \times 10^{-26}$ & -5.4
& tw\tablenotemarkf \\ \tableline
J2051-0827 & $1.2737(5) \times 10^{-20}$ & $<2\times10^{-26}$ & $<$-3.5 & do\tablenotemarkg \\ \tableline
\end{tabular}
\end{center}
\begin{list}{}{}
\tablenotetexta{vs=van Straten et~al. (2001)  Although van Straten et~al. (2001)
did not calculate a limit $\ddot{\nu}$
we have estimated this quantity by
taking the maximum deviation of their timing residuals (100 ns), converting
this to a phase, 
and dividing by the cube of length of their data set, 3.4 y}
\tablenotetextb{la=Lange et~al. (2001)} 
\tablenotetextc{kr=Kramer et~al. (1999) 
See note above\tablenotemarka.
We do the same for this pulsar using 50 $\mu$s and 1700 d.}  
\tablenotetextd{wz=Wolszczan et~al. (2000)}
\tablenotetexte{To calculate an upper limit
we have allowed all the other parameters to vary along
with $\ddot{\nu}$.}
\tablenotetextf{tw=this work}
\tablenotetextg{do=Doroshenko et~al. (2001)}
\end{list}
\end{table*}

The value used for PSR B1937+21 is the best fit to $\ddot{\nu}$
letting all the parameters vary.
Arzoumanian et~al. (1994) were only able to place an upper limit
on $\Delta_8$ for PSR B1855+09 and so are we, but ours is
significantly lower, -6.9, down from -6.  This comes from the best-fit
value of $\ddot{\nu}$ of $ -8 \pm 4 \times 10^{-29}$,
where the uncertainty quoted is that given by TEMPO.
This is significantly smaller than the value
previously obtained by KTR94,
$\ddot{\nu} = -1.0 \pm 0.9 \times 10^{-27}$.  The value
for $\ddot{\nu}$ that we use to calculate $\Delta_8$ is
different from the value one obtains by only allowing
$\ddot{\nu}$ to vary.  With only 1 degree
of freedom, the value is not an appropriate upper limit.   
For J1713+0747 the fit refines nicely to give
$\ddot{\nu} = -2.6 \pm 0.2 \times 10^{-27}$.  

PSR J0437-4715 was discovered by Johnston et~al. (1993).  This 5.8-ms
pulsar is very bright and has been shown to be very good
for timing.  The RMS of the residuals, folded and averaged
at the binary orbital period, is only 35 ns.

Since a third body was discovered around PSR B1257+12, a new upper
limit on $\ddot{\nu} < -1.35\times 10^{-25}$ has been published by
Wolszczan et~al. (2000).  The value is an upper limit since Wolszczan
suggests the distinct possibility
of a fourth body.  While these new measurements do
not change the value of $\Delta_8$
significantly from the value reported by Arzoumanian et~al. (1994), it
is now clear that the value represents an upper limit.

PSR B1534+12 has an updated $\ddot{\nu} < 6\times10^{-28}$
(I.Stairs, private communication)
which yields $\Delta_8 < -5.4$.

PSR B1957+20 is possibly influenced by DM variations, and also by
an instable orbital system, as shown by Applegate \& Shaham (1994) and
Arzoumanian, Fruchter, \& Taylor (1994).  On the basis of the 
relationship between
DM and RMS DM shown by Backer et~al. (1993) we would expect
PSR B1957+20, with a DM of 29, to have an RMS DM of about
0.0002 cm$^{-3}$pc.  We can calculate the expected
influence of these DM variations on the estimation of $\ddot{\nu}$
which was done on single-frequency data at 430 MHz. 
A DM of 0.0002 cm$^{-3}$pc corresponds to a timing residual of
5 $\mu$s.  A 5 $\mu$s trend over the course of
the observations performed by Arzoumanian, Fruchter, \& Taylor (1994) for
example, could produce a false $\ddot{\nu}$ of $\sim 1\times10^{-25}$
given the 5-year observation length.  This $\ddot{\nu}$
corresponds $\Delta_8 = -4.6$, which is the value attributed
to B1957+20 by Arzoumanian et~al. (1994).  In other words, the
timing noise measured by Arzoumanian et~al. (1994) could be
entirely due to unmodeled DM variations.  We therefore
make this point an upper limit and emphasize that the true
value is probably much smaller.

It seems, then, that a significant fraction of MSPs fall below the line
fitted by Arzoumanian et~al. (1994).  Either the relationship is
not well described by a line, or the line has a steeper slope
than was published.  Either way, this bodes well for the
PTA, which relies on submicrosecond accuracy in a handful of
pulsars.

Figure \ref{fig:zaven} is somewhat misleading in that it
makes the $\ddot{\nu}$ measured in B1937+21 look average.
Many of the $\ddot{\nu}$'s measured are from single-frequency
timing, so DM variations may be contributing to what we
are calling `timing noise.'
There is, in fact, no MSP that displays such a convincing,
significant $\ddot{\nu}$.   We measure $\ddot{\nu}$ in B1937+21
to 4 significant digits, where as most other MSP measurements are
upper limits or at best known to 20\%.
It is possible that B1937+21  is
displaying noise whose source is internal to the neutron star crust.
Alternatively, perhaps B1937+21 has
a magnetic field structure
that is evolving with a 20-y timescale, but we see such evolution
at this magnitude in no other object.
As an alternative we consider the possibility that B1937+21 is hosting
a planet.

\section{Planet Around PSR B1937+21?}
\label{sec:planet}

\begin{figure}
\centerline{\psfig{file=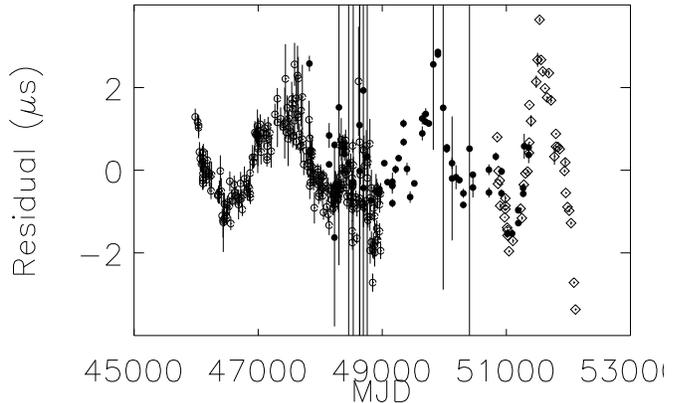,width=8.8cm,clip=}}
\caption{
\label{fig:1937resid_planet}
PSR B1937+21 residuals after fit for a planet.
}
\end{figure}

The residuals shown in part (c) of Figure \ref{fig:master_residual} are
the result of a best-fit model
$\alpha$, $\delta$, $\mu_\alpha$, $\mu_\delta$, $P$, and $\dot{P}$,
and includes daily corrections for DM.
One possible source of their cubic structure is
the existence of a planet of binary period,
$P_b = 17.6 \pm 0.2~y$ which yields an orbital separation of 8 AU,
using Kepler's Law with a central mass of 1.4 \Msolar. 
The residuals after including the planet in the model are shown
in Figure \ref{fig:1937resid_planet}.
A$\sin{i}$
is $27.1 \pm 0.7 \mu$s, yielding a companion mass of
$0.08/\sin{i}$ \Mearth, where $i$ is the inclination of the orbit
to the line of sight.  If real, this planet would be the
smallest known extrasolar planet besides the lunar-mass planet
around PSR B1257+12 (Wolszczan et~al., 2000).  
A planet of similar mass ($0.15/\sin{i}$\Mearth)
has been found around PSR B1257+12
but is much closer to the pulsar (0.20 AU vs 8 AU)
Wolszczan (1994).

\begin{figure}
\centerline{\psfig{file=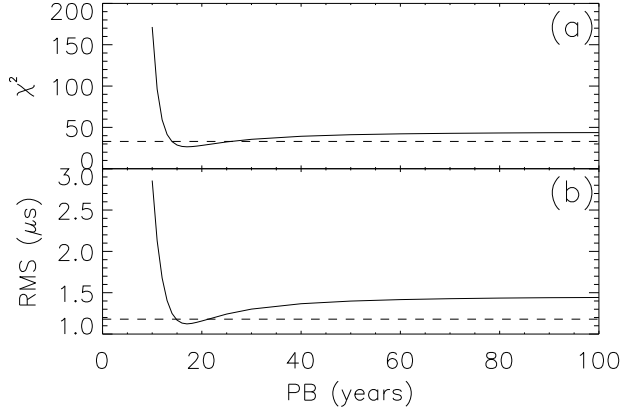,width=8.8cm,clip=}}
\caption[$\chi^2$ vs orbital period for PSR B1937+21]{
\label{fig:xisq_planet}
(a)$\xi^2$ vs orbital period, and (b) RMS vs orbital period
for the model using a planet around B1937+21.  The dotted
line shows the value of the quantity when $\ddot{\nu}$ is
used instead.
}
\end{figure}

The planetary model adds 3 free parameters to the standard
fit, $T_o$, $P_b$, and A$\sin{i}$.  We compare the goodness
of this model to adding the three parameters, 
$\ddot{\nu}$,
$\dddnu$,
and $\ddddnu$.  
The best-fit planetary model fits the data minutely
better than the best $\ddot{\nu}$~through~$\ddddnu$ fit
(RMS of 1.14 $\mu$s vs 1.18 $\mu$s).
The best fit parameters are
$\ddot{\nu} = 1.07(2)\times10^{-26}$s$^{-3}$,
$\dddnu = 1.68(5)\times10^{-34}$s$^{-4}$, 
and $\ddddnu= -1.42(3)\times10^{-42}$s$^{-5}$.
We show in Figure
\ref{fig:xisq_planet} the $\chi^2$ as a function of orbital
period, and RMS as a function of orbital period, both of
which show a minimum near 17 years, which is the length
of our data set.  The dotted line in both parts of the figure
shows the value of the quantity when the fit to $\ddot{\nu}$ is
turned on instead of the addition of the planet.  Only
$\nu$, $\dot{\nu}$, T$_0$, and a$\sin{i}$ were allowed to vary in
the fits.  We did not allow $\alpha$, $\delta$, $\mu_\alpha$ and
$\mu_\beta$ to vary for consistency with the 2-step fitting
process described for B1937+21 in \S \ref{sec:connection}.  

\begin{figure}
\centerline{\psfig{file=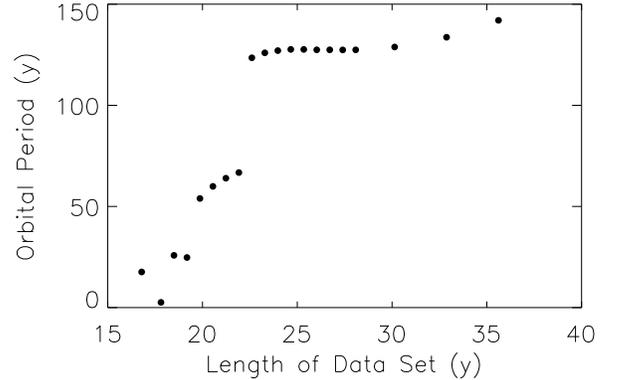,width=8.8cm,clip=}}
\caption[Planetary Orbital Period vs Length of Simulated Data Set]{
\label{fig:plot_simulate}
Best-fit Planetary Orbital Period vs Length of Data Set for simulated
extensions
of PSR B1937+21 data, assuming a constant $\ddot{\nu}$.
}
\end{figure}

It is possible that the residuals you see in
Figure  \ref{fig:master_residual} are the result of some
intrinsic property of the pulsar (see \S \ref{sec:timingnoise}).  
For years this pulsar was
thought to have significant ``timing noise'' causing the
phase to wander as it does (KTR94).
We were concerned that there may always be a well in the
figure of $\chi^2$ vs $P_b$ (e.g. Figure \ref{fig:xisq_planet})
where $P_b$ is roughly equal to the length of the data set.
We tested this possibility by doing the following.  We
simulated an extended data set, N years into the future, by
assuming that $\ddot{\nu}$ is fixed.  We then fit a planet
to each data set and looked at the position of the well
(in other words, the best-fit $P_b$) vs N.  What we found
was that a `planet' will produce a constant $\ddot{\nu}$ up
through the year 2020, which is as far as we tested.  There
is a correlation between orbital period and the length
of the data set, as shown in Figure \ref{fig:plot_simulate}.
Some of the fits are not stable, such as the point
at data span=18 y which yields a 3-y period.  Notice that
in general, the best-fit planetary orbit is much longer than
the data set.  This is easily understood if you consider
that the planetary model needs to imitate the $\ddot{\nu}$ and
can do so by extracting a fraction of an orbit from a
putative planet.  The non-continuous jumps to a higher
$P_b$ shown in Figure \ref{fig:plot_simulate}
are doublings of the orbital
period.  The $\chi^2$ and RMS for these simulated fits
remains very similar to what we observe.  We conclude
that we could be fooled by a constant $\ddot{\nu}$
into supposing the presence of a planet.
However, the previous section, \S \ref{sec:timingnoise}, 
shows that the large and highly significant $\ddot{\nu}$
in B1937+21 is strange relative to the population of
MSPs.  
The only thing that will resolve this is additional $\sim$ 10
years of data, assuming the orbit is about 20 years.

PSR B1937+21 is regarded as having too large a $\dot{P}$ for
being a recycled pulsar, i.e. its large $\dot{P}$ alone implies it
is a 'young' pulsar although it is generally regarded as old
Backer et~al. (1993).  We wondered if perhaps
the unknown existence of an orbiting planet could have been skewing the
measurement of $\dot{P}$ since the discovery of the pulsar.  In
fact, it could not.  Our best fit $\dot{P}$ without the planet, and
without assuming a $\ddot{\nu}$ is slightly smaller, by 1 part
in $\sim10^6$ than our best fit $\dot{P}$ with the planet.

\section{Implications of Planet on Possible Evolutionary Scenarios
of PSR B1937+21}
Millisecond pulsars are thought to be created in supernovae
and spun up by accretion
from a companion star during the red-giant phase of the companion.
Forming planets around such objects is complex and is the
subject of much debate (Wolszczan 1998
and references therein).   In fact, finding planets, such as this
one, in similar millisecond pulsars helps to solve the mystery
in the creation of isolated millisecond pulsars.  At the present,
of the $\sim$70 known millisecond pulsars, 9 are solitary systems,
meaning the presence of a companion has not been detected.  Given
that we suspect these objects have been spun-up by a companion, this
presents an inconsistency.  However, if planets are somehow the remains
of an ablated or otherwise significantly reduced companion, then we
would expect to find similar planets around the other 8 known millisecond
pulsars.  A planet with a 17-y orbit or longer, as we describe would not
have been found in any other system, simply by lack of dataspan.

\section{Conclusion}
\label{sec:conclusion}

We have demonstrated that connection of data sets with sub-$\mu$s
accuracy across multiple telescopes and many years is possible and
yields high precision results for pulsar model parameters.  The
new limit that our data place 
on the energy density in background gravitational radiation,
$ \frac{\rho}{\rho_c} =2 \times 10^{-9}h^{-2} $,
is below that
yielded by the work of KTR94 by more than an order of magnitude.

The data suggest the existence of a small ($< 1$ \Mearth) planet
around B1937+21.  The best-fit orbital period is currently 17.6 y
but that may change with the addition of more data.  We have ruled
out the possibility that the cubic present in B1937+21's residuals
are due to a GW that has geometrically escaped detection in
both J1713+0747 and B1855+09.  We also show that if the cubic is
due to a $\ddot{\nu}$ intrinsic to the pulsar, that B1937+21 is
unique.  We find this alternative therefore unlikely.


\clearpage

\end{document}